\documentclass[amsmath,amssymb,aps,pra,twocolumn]{revtex4-2}
\usepackage{tikz}
\usepackage{graphicx}
\usepackage{amssymb,amsfonts,amsthm}
\usepackage{braket}
\usepackage{hyperref}
\usepackage{color}
\usepackage{soul}
\usepackage{changes}
\usepackage{enumerate,enumitem}
\usepackage{comment}
\usepackage{hyperref}
\usepackage{eqnarray}
\usepackage{bbold}
\usepackage{optidef}
\usepackage{booktabs}

\usepackage{mathtools} 
\mathtoolsset{showonlyrefs} 

\def\vect{\vec{t}}
\def\vecsigma{\vec{\sigma}}
\def\vecs{\vec{s}}
\def\vecr{\vec{r}}

\def\Id{\mathbb{1}}

\def\Tr{\mathrm{Tr}}

\def\opt{\mathrm{opt}}
\def\cl{\mathrm{mp}}

\def\mw{\mathbb{w}}
\def\mR{\mathbb{R}}

\def\opt{\mathrm{opt}}
\def\pa{p^A}
\def\pb{p^B}
\def\overlapADC{O}
\def\teleportationProtocol{P_{a\to B}}
\def\classicalProt{P_\cl}
\def\classicalProtOpt{P_\cl^\opt}
\def\SQTBDProtocol{P_{\text{Bd}}}
\def\SQTADProtocol{P_{\text{ad}}}
\def\initialState{\rho^a}
\def\pdf{f}
\def\PDFClassicalProt{\pdf^{\cl}}
\def\blochVectorOutputStates{\vect_{j|\vect}}
\def\PDFClassicalProtOpt{\pdf^\cl_{\opt}}
\def\ProbClassical{P^\cl}
\def\ProbClassicalOpt{P^\opt}
\def\ProbPhaseFlips{P^\pf}
\def\avPhaseFlips{\braket{F}^\pf}
\def\varPhaseFlips{\text{Var}^\pf(F)}   
\def\varProtClassOpt{\text{Var}_{\opt}^\cl(F)}
\def\PDFBellDiag{\pdf^{\text{Bd}}}
\def\PDFBellDiagPhaseFlip{\pdf^{\text{pf}}}
\def\PDFADC{\pdf^{\text{ad}}}
\def\outputStateJDataVect{\rho_{j|\vec{t}}}
\def\ProbJDadoVect{p_{j|\vec{t}}}
\def\pf{\text{pf}}

\def\diag{\text{diag}}

\def\corrmatElement{w}
\def\corrmat{\mathbb{w}}
\def\depol{\text{d}}
\def\importancePrior{\text{W}}
\newcommand{\importancePriorBeta}[2]{\importancePrior_{ #1 , #2 } }

\def\adcFunc{c_{\text{ad}}}
\def\phaseflipFunc{c_{\text{pf}}}

\DeclareMathOperator*{\argmax}{argmax}

\theoremstyle{theorem} 
\newtheorem{result}{Proposition}

\theoremstyle{theorem} 
\newtheorem{corollary}{Corollary}

\newtheorem{definition}{Defintion}

\usepackage{xcolor}

\begin{document}


\title{Quality assessment of quantum teleportation through the distribution of fidelity}

\author{D. G. Bussandri}
\email{diegogaston.bussandri@uva.es (corresponding author)}
\affiliation{Departamento de F\'{\i}sica Te\'orica, At\'omica y Optica and Laboratory for Disruptive \\ Interdisciplinary Science (LaDIS), Universidad de Valladolid, 47011 Valladolid, Spain}

\author{G. M. Bosyk}
\email{gbosyk@icc.fcen.uba.ar}
\affiliation{CONICET-Universidad de Buenos Aires, \\Instituto de Ciencias de la Computaci\'on (ICC), Buenos Aires, Argentina}

\author{P. Crespo Del Amo}%
\email{}
\affiliation{Departamento de F\'{\i}sica Te\'orica, At\'omica y Optica, Universidad de Valladolid, 47011 Valladolid, Spain}


\author{K. Życzkowski}%
\email{}
\affiliation{
Faculty of Physics, Astronomy and Applied Computer Science, Jagiellonian University, ul. Łojasiewicza 11, 30-348 Kraków, Poland\\
Center for Theoretical Physics, Polish Academy of Sciences,
Al. Lotników 32/46, 02-668 Warszawa, Poland
}


\begin{abstract}

In this work, we introduce a comprehensive statistical framework for assessing single-qubit quantum teleportation performance beyond the conventional average‑fidelity benchmark. 
At first, we derive a closed-form expression for the full probability density function of actual teleportation fidelities and apply it to both classical 'measure-and-prepare' schemes and standard quantum teleportation, considering two relevant noise models: Bell-diagonal resource states and local amplitude-damping channels. 
These results reveal that protocols with identical average fidelities can exhibit markedly different statistical behaviors, and that relying solely on average fidelity can mask inherent asymmetries introduced by local noise, potentially leading to spurious conclusions of symmetry.
Secondly, we introduce a certification method based on prior importance functions (e.g., Beta distributions), which unifies moment‑based criteria and threshold‑based success probabilities into a single figure of merit. 
Applying this framework, we show that certifying high‑fidelity teleportation requires increasingly stronger entanglement or non‑locality, and we clarify that the so‑called ``fighting noise with noise'' effect arises from the chosen prior importance function rather than representing a genuine advantage. 
Our approach thus provides versatile tools for tailored, application‑specific teleportation benchmarks.

\end{abstract}

\maketitle

\section{Introduction}

Quantum teleportation \cite{Bennett1993} is a fundamental protocol in quantum information science, with applications in quantum communication, quantum computing, and emerging quantum technologies \cite{Pirandola2015,Hu2023}. 

In its ideal form, an arbitrary quantum state is deterministically transmitted from a sender (Alice) to a distant receiver (Bob) by employing a maximally entangled pair and two bits of classical communication, yielding perfect identity between input and output. 
In practice, however, environmental noise on the entangled resource and implementation imperfections degrade the protocol's performance, so that Bob’s received state may deviate from Alice’s original \cite{Oh2002,Badziag2000,Bandyopadhyay2000,Knoll2014,Fortes2015,Im2021,Bussandri2024}.

To date, most teleportation benchmarks rely on the \textit{average fidelity}~\cite{Pirandola2015}. 
Although this single‐number metric is convenient, it may mask crucial statistical features. In particular, the average fidelity represents the actual fidelity value averaged over an infinite number of realizations of the protocol with maximally random input states, and it also coincides with the average value of the \textit{channel} or \textit{prior} fidelity over the set of input random states. In the first case, the random variable to be averaged is the actual fidelity value, while in the second case, it is the channel fidelity. These random variables coincide in average but, in general, do not have the same statistical distribution. Furthermore, in any case, taking solely the average value disguises potential different statistical behaviors: two protocols may exhibit identical average fidelities yet differ dramatically in variance and tail behavior of their fidelity distributions.

Several studies within the context of teleportation protocols have begun to explore higher-order moments of the channel (or prior) fidelity, focusing particularly on its deviation rather than the full output distribution \cite{Bang2018,Ghosal2020a,Ghosal2020b,Roy2020}. On the other hand, full statistical fidelity distributions were studied in the context of quantum state transfer protocols and operations or channels applications, see Refs.~\cite{Lorenzo2025,Chełstowski2023}.

Consequently, a comprehensive, unified framework for characterizing quantum teleportation performance under noise — and a certification method based on the full fidelity distribution — remains elusive.

In this work, we address two significant gaps for the case of single-qubit teleportation. This protocol remains a cornerstone of current quantum communication experiments, as evidenced by recent advances in metropolitan, network, and satellite-based implementations; see, for example, \cite{valivarthi_quantum_2016,sun_quantum_2016,ren_ground-to-satellite_2017,valivarthi_teleportation_2020,zuo_overcoming_2021,rakonjac_transmission_2023}.

First, we obtain a simple expression for the full continuous probability distribution (i.e. the probability density function) of the actual fidelity values, and then for protocols such as classical and standard teleportation, we obtain closed-formulas revealing their main statistical features that average fidelity may overlook. Within standard quantum teleportation, we considered two noise models: Bell-diagonal states as a resource of the protocol and an ideal Bell state affected by two amplitude-damping channels. Notably, in this last case, we show that although the average fidelity remains unchanged under local damping, the full distribution does not, revealing spurious symmetries that would otherwise go unnoticed.
In other words, average fidelity can lead to symmetries that are not inherent to the protocol, potentially leading to erroneous conclusions about its true performance and symmetry.

Second, we introduce a certification framework based on \textit{prior importance functions}, which unifies a broad class of certification criteria: from distribution moments (average, variance) to threshold-based success probabilities. 
These functions recognize that certain outcomes (e.g., unit fidelity) may be prioritized, while others (e.g., low-fidelity states) require avoidance.
This approach enables tailored assessments of teleportation schemes according to application-specific requirements.

Applying this framework to Bell-diagonal resource states, we find that certifying high-fidelity teleportation demands resource states with increasingly strong entanglement (or non-locality), suggesting an operational link between teleportation certification and non-locality.

Finally, we revisit the ``fighting noise with noise'' phenomenon \cite{Badziag2000,Bandyopadhyay2000,Knoll2014,Fortes2015}, demonstrating that its apparent advantage depends critically on the chosen importance function and vanishes in the high-fidelity regime, suggesting it may be an artifact of certification rather than a genuine physical effect.

\textit{Structure of the article.} Section \ref{sec:Teleportation_protocols} provides a comprehensive review of teleportation protocol fundamentals. We begin in Section \ref{sec:IntroDistributions} by introducing state-fidelity-based figures of merit for evaluating teleportation performance. Section \ref{sec:measurementPrepare} presents classical ``measure-and-prepare" schemes, while Section \ref{sec:standard} examines standard quantum teleportation implemented with imperfect two-qubit resources, focusing on two distinct noise models: local unital channels and local amplitude-damping channels.
Section \ref{sec:Teleportation_fidelity_statistics} presents our derivation of the fidelity probability density function for the general case, yielding closed-form analytical expressions for both classical protocols and standard quantum teleportation. These formulas apply specifically to Bell-diagonal resource states and resources subject to local amplitude-damping noise.
Section \ref{sec:prior_importance_certification} develops a novel certification framework based on prior importance functions. We analyze the performance of Bell-diagonal states and amplitude-damped resources under various Beta-distribution priors in Sections \ref{sec:certification_Bell_diagonal} and \ref{sec:certification_ADC}, respectively.
Section \ref{sec:conclusions} concludes with a summary of our key findings and explores their implications for developing robust teleportation benchmarks.

\section{Teleportation protocols}\label{sec:Teleportation_protocols}

Consider a set of initial states $\{\initialState_i\}_i$ of a quantum system $a$ prepared according to a certain probability distribution $p=\{p_i\}_i$. 
A teleportation protocol $\teleportationProtocol$ is intended to transfer states $\initialState_i$ to another quantum system $B$. In practice, $\teleportationProtocol$ delivers output states $\rho^B_{j|i}$ with probability $p_{j|i}$, where the index $j$ enumerates the possible outcomes of the protocol given the input $\initialState_i$. In this context, indices $i$ can be discrete or continuous, whereas indices $j$ shall always be discrete. By convention, we will refer to Alice as the sender and Bob as the receiver.

\subsection{Assessing performance}
\label{sec:IntroDistributions}


Distance-based benchmark measures fundamentally rely on quantifying the difference between initial and output states through a chosen distance measure~\cite{Massar1995,Popescu1994,Bussandri2024}. In teleportation or state transfer protocols, the Uhlmann-Jozsa Fidelity~\cite{Uhlmann1976,Jozsa1994}, defined as 
$F(\rho,\sigma) = \left(\Tr(\sqrt{\sqrt{\rho}\, \sigma \sqrt{\rho}})\right)^2$, is the predominant quantity for assessing state distinguishability~\cite{Pirandola2015}. This work will focus on this particular measure.

Fixed the distance, different quantities can be considered to assess the performance of teleportation protocols. For a given initial state $\initialState_i$, the \textit{prior fidelity} 
\begin{align}\label{eq:PriorFidelity}
   \bar{F}_i=\sum_j p_{j|i} F(\initialState_i , \rho^B_{j|i}),
\end{align}
provides a measure of the expected fidelity across all possible outcomes (the index $j$ runs across all possible outcomes)~\cite{Vidal1999}. In contrast, the \textit{channel fidelity}~\cite{Chełstowski2023,Lorenzo2025},
\begin{align}\label{eq:ChannelFidelity}
   F^{\text{ch}}_i(\initialState_i)= F(\initialState_i , \sum_j p_{j|i} \rho^B_{j|i}),
\end{align}
is particularly relevant when assessing quantum state transfer channels, often referred to as \textit{transport}, where no post-selection is involved. The resulting output state is then directly represented by $\Phi_P(\initialState_i)=\sum_j p_{j|i} \rho^B_{j|i}$, which reflects the channel's transformation of the initial state. Additionally, for pure input states (which is the main assumption in most theoretical and experimental teleportation studies), the prior fidelity, Eq.~\eqref{eq:PriorFidelity}, results equal to the channel fidelity, Eq.~\eqref{eq:ChannelFidelity}, because of fidelity's linearity for pure inputs. 

The \textit{average fidelity} $\left< \bar{F}_i \right>_i$, over all (pure and random) input states $\initialState_i$, is the usual measure to assess teleportation protocols' performance, playing a central role in the most famous criteria for \textit{successful teleportation}: a given teleportation protocol $\teleportationProtocol$ is certified as successful if and only if its average fidelity is greater than  $2/3$, the highest average fidelity achievable with classical protocols~\cite{Pirandola2015}. 

As we are interested in teleportation protocols involving classical communication and post-selection, and because we aim to understand the fidelity behavior when the protocol is performed, we shall focus on the statistics of the actual fidelity,
\begin{align}\label{eq:PosteriorFidelity}
    F(\initialState_i,\rho^B_{j|i}),
\end{align} 
which, according to Bayes' theorem, is ruled by the joint probability distribution $p_{ij}=p_ip_{j|i}$. 


In the following sections, we will employ a continuous set of initial states distributed according to the Haar measure. Using the Bloch parametrization, each initial state can be written as
\begin{equation}\label{eq:InitialState}
	\initialState_{\vec{t}} = \frac{1}{2} \left(\Id + \vect \cdot \vecsigma \right),
\end{equation}
and an arbitrary teleportation protocol $\teleportationProtocol(\initialState)$ is determined by the ensemble $\{\ProbJDadoVect,\outputStateJDataVect^B\}_j$, where $\vec{t}\in S$ belongs to the Bloch sphere $S$. Accordingly, for a fixed initial state $\initialState_{\vec{t}}$, the corresponding output states of the protocol are given by $\outputStateJDataVect=\frac{1}{2}(\mathbb1 + \blochVectorOutputStates\cdot\vec\sigma)$
and the fidelity takes the value 
\begin{align}\label{eq:PosteriorFidelityCont}
F_j(\vec{t})=F(\initialState_{\vec{t}},\outputStateJDataVect^B)=\frac{1}{2}(1+\vec{t}\cdot\vec{t}_{j|\vec{t}}),
\end{align}
with probability $\ProbJDadoVect$. Note that the index $j$ labels the possible outputs of the teleportation protocol.

The joint probability density for obtaining a fidelity 
$F(\initialState_{\vec{t}},\outputStateJDataVect^B)$, with $\vec{t}$ in the Bloch 
sphere solid angle $d\Omega$, is:
\begin{align}\label{eq:JointProb}
    \frac{1}{4\pi}d\Omega \ProbJDadoVect.
\end{align}
This expression directly results from the total probability law in its continuous form: the term $\frac{1}{4\pi}d\Omega$ gives the probability of selecting $\vec{t}$, 
while $\ProbJDadoVect$ specifies the conditional probability of the $j$th outcome (given $\vec{t}$).

\subsection{Classical Teleportation Protocols}
\label{sec:measurementPrepare}

Classical teleportation uses no quantum resources: Alice and Bob are restricted to sharing information through a classical channel; thus, they are limited to performing a measurement-prepare protocol. The outline of this procedure is as follows.

Alice performs a POVM measurement given by $\mathcal{E}=\{E_j\}_j$ over the quantum system $a$, which we assume is occupying one of the initial states in Eq.~\eqref{eq:InitialState}. Using the Bloch parametrization, the operators $\{E_j\}_j$ can be written in terms of normalization constants $c_j$ and Bloch vectors $\vecs_j$~\cite{Vidal1999}:
\begin{equation}
	\label{eq:POVM_def}
	E_j = \frac{c^2_j}{2} \left(\Id + \vecs_j \cdot \vecsigma \right),
\end{equation}
which satisfy:
\begin{equation}
	\label{eq:POVM_conditions}
	\sum^n_{j=1} c^2_j = 2 \ \text{and} \ \sum^n_{j=1} c^2_j \vecs_j= \vec{0}.
\end{equation}
If Alice obtains the $j$-th measurement result, she uses the classical channel to communicate it to Bob, who prepares the output of this classical protocol according to a conditional strategy:
\begin{equation}\label{eq:ClassicalOutputs}
	\rho^{\cl,B}_{j|\vec{t}} = \frac{1}{2} \left(\Id + \vecr_j \cdot \vecsigma \right),
\end{equation}
The $j$-th output occurs with probability,
\begin{equation}
    \label{eq:probabilidades_clasicas}
	\ProbJDadoVect^\cl = \frac{c_j^2}{2} \left(1 + \vect \cdot \vecs_j\right),
\end{equation}
and thus $\ProbJDadoVect^\cl$ stands also for the probability of preparing $\rho^{\cl,B}_{j|\vec{t}}$.

In summary, every classical protocol is defined by a measurement, specified by elements $\{c_j,\vec{s}_j\}_j$, and a preparation determined by $\{\vec{r}_j\}_j$. Among these classical protocols, we shall refer to as \textit{optimal} those maximizing the average fidelity, i.e. those measurements ($\{c_j,\vec{s}_j\}_j$) and preparations ($\{\vec{r}_j\}_j$) such that $\braket{F}^\cl_\opt=2/3$, see Ref.\cite{Massar1995,Vidal1999}. Following these seminal references, the condition for a protocol to be optimal is $\vec{r}_j\cdot\vec{s}_j=1$.

\subsection{Standard quantum teleportation protocol}\label{sec:standard}

Standard quantum teleportation involves teleportation protocols implemented over an imperfect resource state $\rho^{AB}$. This situation arises, for example, when $\rho^{AB}$ is directly produced by an imperfect source or when an ideal resource state is affected by environmental noises. By the usual definition of this protocol, the resource state is taken as an arbitrary two-qubit quantum state~\cite{Fano1983},
\begin{small}
    \begin{align}\label{eq:ResourceStateAB}
\! \! \!\!\!\!\!\!\! \rho^{AB}\!=\!\frac{1}{4}\!\left(\mathbb{1}\!\otimes\!\mathbb{1}\! +\! \vec{r}^A\!\cdot\!\vec{\sigma}\! \otimes\!\mathbb{1}\!+\!\mathbb{1}\!\otimes\!\vec{r}^B\!\cdot\!\vec{\sigma}\!+\!\!\!\sum_{ij=1}^3 [\mathbb{w}]_{ij}\sigma_i\!\otimes\!\sigma_j \right)\!,
\end{align}
\end{small}

\noindent where the correlation matrix reads, $\corrmat = \Tr\left(\rho^{AB} \,\vec{\sigma} \otimes \vec{\sigma} \right)$. The previous parametrization is known as \textit{Fano form}~\cite{Fano1983}.

The protocol procedure is fundamentally the same as in ideal quantum teleportation: In the first step, Alice performs a projection onto the Bell basis (also referred to as Bell measurement), $\{E^{aA}_i=\ket{\Phi_i}\!\bra{\Phi_i}\}_{i=1}^4$,
\begin{align}
  \ket{\Phi_{1}}&\!=\!\frac{1}{\sqrt{2}}\!\left(\ket{00}\!+\!\ket{11} \right), &  \ket{\Phi_{3}}&\!=\! \frac{1}{\sqrt{2}}\!\left(\ket{01}\!+\!\ket{10} \right), \label{eq:Bellstates} \\ 
\ket{\Phi_{2}}&\!=\! \frac{1}{\sqrt{2}}\!\left(\ket{00}\!-\!\ket{11} \right), & \ket{\Phi_{4}}&\!=\! \frac{1}{\sqrt{2}}\!\left(\ket{01}\!-\!\ket{10} \right).  \label{eq:Bellstates1} 
\end{align}
Bell states have maximally mixed reduced states and, thus, in their Fano form, are completely characterized by their correlation matrices  $\mathbb{w}_i = \text{Tr} \left(\ket{\Phi_i}\bra{\Phi_i} \vec \sigma \otimes \vec \sigma \right)$, with $i=1,\dots, 4$; specifically:
\begin{align}\label{eq:BellstatesCorrMat}
	\mathbb{w}_1&=\text{diag}(1,-1,1), &  \mathbb{w}_3&=\text{diag}(1,1,-1),\\
	\mathbb{w}_2&=\text{diag}(-1,1,1), & \mathbb{w}_4&=\text{diag}(-1,-1,-1). \label{eq:BellstatesCorrMat1}
\end{align}
After Alice has obtained the $j$-th result, with probability 
\begin{align}\label{eq:probSQT}
    \ProbJDadoVect=\frac{1+ \vec{t}\cdot (\mathbb{w}_j\vec{r}^A)}{4},
\end{align}
the resulting reduced state in system $B$ is:
\begin{align}\label{eq:CondState}
    \sigma^{B}_{j|\vec{t}} &= \frac{1}{2} \left(\Id + \vec{s}^{\,B}_{j|\vec{t}} \cdot \vecsigma \right), \ \text{where} \\
    \vec{s}^{\,B}_{j|\vec{t}}&=\frac{\vec{r}^B+(\mathbb{w}_j \corrmat)^\intercal \vec{t}}{4 \ProbJDadoVect}.  \label{eq:conditionalVect}
\end{align}
Here, $\vec{r}^A$ and $\vec{r}^B$ are the Bloch vectors of the reduced states, and $\corrmat$ is the correlation matrix corresponding to the resource state $\rho_{AB}$ [see Eq.~\eqref{eq:ResourceStateAB}]. The matrices $\{\corrmat_i\}_{i=1}^4$, in turn, are the correlation matrices of the Bell states, defining Alice's Bell measurement.

In the next step, Alice uses a classical channel to send its result to Bob, who applies a unitary operation $U_j$:
\begin{equation}
\label{eq:outputState}
	\rho^{B}_{j|\vec{t}} =U_j \sigma^{B}_{j|\vec{t}} \, U_j^\dagger=\frac{1}{2}(\mathbb{1}+\blochVectorOutputStates \cdot \vec{\sigma}), \ \text{and } \vec{t}_{j|\vec{t}}=\mR_j\vec{s}^{\,B}_{j|\vec{t}}.
\end{equation}
This conditional Bob operation, $U_j$, is chosen to optimize the average fidelity. In standard quantum teleportation, these operations are~\cite{Bussandri2024}:
\begin{align}
    \mR_j^{\text{opt}}\!&=\!\mw_j \mathbb{O}_1\mw_{l(\corrmat_d)}\mathbb{O}_2^\intercal,\! &
    l(\corrmat_d)\!&=\!\argmax_i \! \left\{\!1\!+\!\Tr[\mw_i\corrmat_d] \right\},
\end{align}
where the orthogonal rotation matrices $\mathbb{O}_i\in \text{SO}(3)$ are determined by the singular value decomposition of the correlation matrix $\corrmat$, specifically, $\corrmat=\mathbb{O}_1 \corrmat_d \mathbb{O}_2^\intercal$ (note that $\corrmat_d$ can have negative elements)~\cite{Horodecki1996a}. This outlines the standard quantum teleportation protocol.

\subsubsection{Noise models}\label{sec:noise_models}

In the following section, we will analyze two kinds of resource states. The first are \textit{Bell-diagonal} states, i.e. states being diagonal in the Bell basis. This implies maximally reduced states and, in the Fano form~Eq.~\eqref{eq:ResourceStateAB}, the correlation matrix is diagonal:
\begin{align}\label{eq:BellDiagonalStates}
    \vec{r}^A&=\vec{r}^B=\vec{0}, &
    \corrmat&=\text{diag}\{\corrmatElement_1,\corrmatElement_2,\corrmatElement_3\}.
\end{align}

This set of states can also be written as local noises, represented by unital qubit channels, affecting a perfect Bell state. Specifically, when two quantum channels $\mathcal{E}$ and $\mathcal{F}$ with affine decompositions $(A_{\mathcal{E}},\vec{b}_{\mathcal{E}})$ and $(A_{\mathcal{F}},\vec{b}_{\mathcal{F}})$, respectively, act on a Bell state $\left|\Phi_k\right>\!\left<\Phi_k\right|$, the resulting resource state is given by~\cite{Bussandri2024}:
\begin{align}\label{eq:noiseschannels}
     &\mathcal{E} \otimes \mathcal{F}\left(\left|\Phi_k\right>\!\left<\Phi_k\right|\right) = \frac{1}{4}\left[\mathbb{1}_4+\vec{b}_{\mathcal{E}} \cdot \vec{\sigma} \otimes \mathbb{1}+\mathbb{1} \otimes \vec{b}_{\mathcal{F}} \cdot \vec{\sigma}\right.\nonumber\\&\left.+\sum_{\alpha \beta = 1}^3\left(b_{\mathcal{E}}^\alpha b_{\mathcal{F}}^\beta+\sum_{i=1}^3\mathbb{w}_k^{ii} A_{\mathcal{E}}^{\alpha i} A_{\mathcal{F}}^{\beta i}\right) \sigma_\alpha \otimes \sigma_\beta\right].
\end{align}
From Eq.~\eqref{eq:noiseschannels} we can derive Eqs.~\eqref{eq:phaseflipfunc}, \eqref{eq:depol_resource_state_mat_corr}, \eqref{eq:TWOADCReducedStates}, and~\eqref{eq:TWOADCCOrrMat}.
In particular, if each local noise is a phase flip channel, given by $\vect_\pf=\eta_{\pf}\vect$ with $\eta_\pf=\text{diag}\{1-p,1-p,1\}$, the resulting resource state becomes Bell-diagonal with correlation matrix given by 
\begin{align}
    \corrmat_\pf&=\diag\{1-\phaseflipFunc,\phaseflipFunc-1,1\} \text{ with }  \label{eq:phaseflipfunc} \\ 
    \phaseflipFunc&=(\pa+\pb)-\pa\pb. \label{eq:phaseflipfunc2}
\end{align}
Here, $\pa$ and $\pb$ are the corresponding noise parameters to subsystems $A$ and $B$, respectively ($\phaseflipFunc\in[0,1]$). 


On the other hand, if the local noises are depolarizing channels, characterized by $\vect_\depol=\eta_{\depol}\vect$ with $\eta_{\depol}=(1-p)\mathbb{1}$, the resource state results in a Werner state with correlation matrix 
\begin{align}\label{eq:depol_resource_state_mat_corr}
    \corrmat_\depol=(1-\pa)(1-\pb)\mw_1.
\end{align}


In the second case, we will consider a noise model leading to a non-Bell-diagonal resource: two local amplitude-damping channels acting over systems $A$ and $B$, which results in
\begin{align}
\vec{r}^A&=\pa\hat{k}, \ \vec{r}^B=\pb\hat{k}, \label{eq:TWOADCReducedStates} \\
\corrmat_{\text{ad}}&=\mw_1\text{diag}\{\adcFunc,\adcFunc,\adcFunc^2\}+\pa\pb\hat{k}\hat{k}^\intercal,\label{eq:TWOADCCOrrMat}\\
\adcFunc&=\sqrt{\!(1\!-\!\pa)\!(1\!-\!\pb)}. \nonumber
\end{align}

\section{Teleportation fidelity statistics}\label{sec:Teleportation_fidelity_statistics}

Let us consider a teleportation protocol $\teleportationProtocol$ described by the ensemble $\{\ProbJDadoVect,\outputStateJDataVect\}_j$ given an initial state $\rho_{\vec{t}}^a$ distributed according to the --continuous-- Haar measure. Following Sec.~\ref{sec:IntroDistributions}, the teleportation fidelity statistic is characterized by the statistical behavior of $F_j(\vec{t})$, Eqs.~\eqref{eq:PosteriorFidelityCont}, ruled by the continuous joint probability distribution of getting $j$ given $\vect$, Eq. \eqref{eq:JointProb}. The actual fidelity probability density function $\pdf(F)$ can be calculated employing the following general result, derived in Appendix~\ref{app:Distr}.

\begin{result}\label{res:fidelity_prob_distr}
If $F_j(\vect)=F(\initialState_{\vec{t}},\outputStateJDataVect^B)$, see Eqs.~\eqref{eq:PosteriorFidelityCont}, is $C^\infty(S)$, with $S$ the Bloch sphere, the probability density function of the fidelity can be written as, 
\begin{align} \label{eq:ProbDistrFid}
   \pdf(F) =\sum_j \frac{1}{4\pi}\int d \mathcal S_j \  \frac{p_{j|\vec{x}}}{\left|\nabla F_j(\vec{x})\right|},
\end{align}
where $d \mathcal S_j$ is the corresponding Euclidean surface measure corresponding to $\mathcal S_j=\{\vec{x}\in S; F_j(\vec{x})=F\}$, and $F\in [0,1]$. The index $j$ labels the possible outputs of the protocol. 
\end{result}

\subsection{Measurement-prepare protocol}\label{sec:Classical_teleportation}

As we introduce in Sec. \ref{sec:measurementPrepare}, a measurement-prepare protocol is given by the ensemble $\{\ProbJDadoVect,\outputStateJDataVect^{\cl,B}\}_j$, where the probabilities $\ProbJDadoVect$ are in Eq.~\eqref{eq:probabilidades_clasicas} and $\outputStateJDataVect^{\cl,B}$ in Eq.~\eqref{eq:ClassicalOutputs}. The following result, demonstrated in Appendix~\ref {app:ClassicalDistr}, specifies the probability density function corresponding to these classical protocols.

\begin{result}\label{res:probDistrClassProt}
    The probability density function of the fidelity values produced by the measurement-prepare protocol $\{\ProbJDadoVect,\outputStateJDataVect^{\cl,B}\}_j$ is,
\begin{align}\label{eq:probDistrClassProt}
    \PDFClassicalProt(F)&=\sum_{j} \PDFClassicalProt_{j}(F), \ \text{ with }  \\
    \PDFClassicalProt_{j}(F)&=
        \begin{cases}
         \!\frac{c_j^2}{2r_j}\!\left(\!1\!-\!\frac{\vec{r}_j\cdot\vec{s}_j}{r_j ^2}\right)\!+\!F\frac{c_j^2\vec{r}_j\cdot\vec{s}_j}{r_j ^3}, & \!F\in[\frac{1-r_j}{2},\frac{1+r_j}{2}] \\
        0, & \!F\not\in [\frac{1-r_j}{2},\frac{1+r_j}{2}],
        \end{cases}
\nonumber
\end{align}
where $r_j=|\vec{r}_j|$ and $c_j$ is defined through Eqs.~\eqref{eq:POVM_def} and~\eqref{eq:POVM_conditions}.
\end{result}

The probability of obtaining a value $F$ for the fidelity in the interval $[F_a,F_b]$ is then,
\begin{align}
    &\ProbClassical(F\in[F_a,F_b])\!= \int_{F_a}^{F_b} \PDFClassicalProt(F) dF \\
    &=\!\sum_{j}\!\frac{c_j^2(\!F_{b,j}\!-\!F_{a,j}\!)}{2r_j}\!\left[1\!-\!\frac{\vec{r}_j\cdot\vec{s}_j}{r_j^2}(1\!-\!F_{a,j}\!-\!F_{b,j})\right]
\end{align}
with 
$F_{a,j}=\max \{F_a, \frac{1-r_j}{2}\}$ and $F_{b,j}=\min \{F_b, \frac{1+r_j}{2}\}$. Additionally, the average fidelity is given by:
\begin{align}
    \braket{F}^\cl &=\int_{0}^{1} \PDFClassicalProt(F) \ F dF \\
    &= \frac{1}{2}\left(1+\frac{1}{3}\sum_j \frac{c_j^2}{2}\vec{r}_j\cdot\vec{s}_j\right).
\end{align}
The previous expression coincides with the one obtained in Ref.~\cite{Vidal1999}. 

As we can see from Eq.~\eqref{eq:probDistrClassProt}, the fidelity produced by classical protocols has a linear behavior: Its probability density function is composed of the sum of linear functions with support $[ (1-r_j)/2, (1+r_j)/2 ]$, therefore centered at $F=1/2$, and slopes $\frac{c_j^2\vec{r}_j\cdot\vec{s}_j}{r_j ^3}$.

The probability density function of the optimal protocol, characterized by $\vec{r}_j\cdot\vec{s}_j=1$ and leading to the average fidelity of the measurement-prepare protocol $\braket{F}^\cl_\opt = 2/3$~\cite{Massar1995,Vidal1999}, is also a particular case of Result~\ref{res:probDistrClassProt}. This leads to the following corollary:
\begin{corollary}
    The probability density function of optimal measurement-prepare protocols (i.e. $\vec{r}_j\cdot\vec{s}_j=1$ for all $j$), is,
\begin{align}\label{eq:probDistrClassProtOpt}
    \PDFClassicalProtOpt(F) = 2F.
\end{align}
\end{corollary}

Thus, the optimal classical protocol produces a fidelity value in the range $F\in[F_a,F_b]$ with probability,
\begin{align}
    \ProbClassicalOpt(F\in[F_a,F_b])\!=\int_{F_a}^{F_b} \PDFClassicalProtOpt(F) dF=F_b^2-F_a^2,
\end{align}
and the average and variance of the protocol are
\begin{align}
   \braket{F}^\cl_\opt&= 2/3,\label{eq_classical_av_fid}\\
    \varProtClassOpt &= 1/18.
\end{align}
The previous values are calculated employing the probability density function in Eq.~\eqref{eq:probDistrClassProtOpt}.

\subsection{Standard quantum teleportation}\label{sec:Standard_quantum_teleportation}

\subsubsection{Bell-Diagonal States}
In this case, the reduced states are maximally mixed and the correlation matrix is diagonal, see Eq.~\eqref{eq:BellDiagonalStates}. This resource state leads to a \textit{deterministic} protocol in the following sense: The output states $\outputStateJDataVect$ are equal for all of Alice's measurement outcomes, 
\begin{align}
\blochVectorOutputStates=\mathbb{R}_j^{\text{opt}}\mathbb{w}_j\corrmat_d \vec{t}= \mw_{l(\corrmat_d)} \corrmat_d \vec{t},
\end{align}
and occurs with the same probability, $\ProbJDadoVect=1/4$. Thus, the fidelity results in the following quadratic form,
\begin{align}
    F_j(\vec{t}) = \frac{1}{2}\left[1+\vect\cdot \mw_{l(\corrmat_d)} \corrmat_d \vec{t} \ \right].\label{eq_quadratic_fid}
\end{align}
The fidelity probability density function implied by Eq.~\eqref{eq_quadratic_fid} (and by an arbitrary quadratic form) is given in the following result derived in Appendix~\ref{app:QuatraticFormDistr}.

\begin{result}\label{res:BD_prob_distr}
The probability density function of the fidelity for the standard quantum teleportation protocol with a Bell-diagonal resource $\SQTBDProtocol$ is given by the complete elliptic integral of the first kind $K(\kappa)$,
\begin{align}\label{eq:distrProbBellDiagonal}
    \PDFBellDiag(F)=\begin{cases}
        \frac{2K\left[\frac{(a_1-a_2)(2F-1-a_3)}{(a_2-a_3)(a_1-2F+1)}\right]}{\pi\sqrt{(a_2-a_3)(1+a_1-2F)}}, & F\in\left[\frac{1+a_3}{2},\frac{1+a_2}{2}\right], \\
        \frac{2K\left[\frac{(a_2-a_3)(a_1-2F+1)}{(a_1-a_3)(a_2-2F+1)}\right]}{\pi\sqrt{(a_1-a_3)(2F-1-a_2)}}, & F\in\left(\frac{1+a_2}{2},\frac{1+a_1}{2}\right], \\
        0, & F\not\in[\frac{1+a_3}{2},\frac{1+a_1}{2}],
    \end{cases}
\end{align}
where the numbers $a_i$ are the elements $[\mathbb{w}_{l(\corrmat_d)}\corrmat_d]_i$ ordered in decreasing sequence (i.e. $a_1$ is the maximum of $\{[\mathbb{w}_{l(\corrmat_d)}\corrmat_d]_i\}$). 
\end{result}

To illustrate the statistical behavior of fidelity within this type of resource state, we analyze the effects of local phase flip and depolarizing channels, representing local noises affecting a perfect Bell state $\ket{\Phi_1}$. Explicit expressions are included in see Sec.~\ref{sec:noise_models}. 


\begin{figure}[htb]
    \centering
    \includegraphics[width=\linewidth]{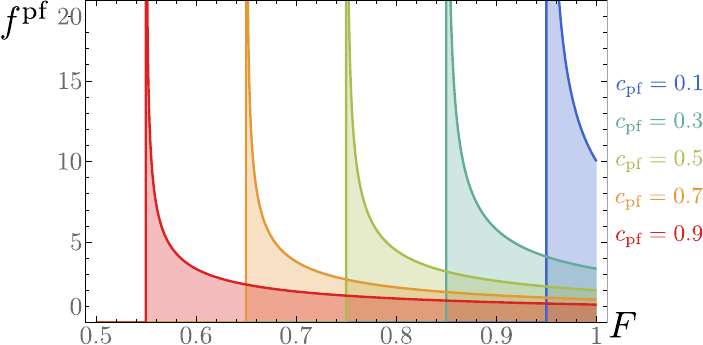}
    \caption{Plots of the probability density function $\PDFBellDiagPhaseFlip(F)$, Eq.~\eqref{eq:distrProbBellDiagonal_phase_flip}, for different values of $\phaseflipFunc$, Eq.~\eqref{eq:phaseflipfunc}. This case corresponds to the standard quantum teleportation protocol for an ideal resource (Bell state $\ket{\Phi_1}$) affected by two local phase flip noises with parameters $p_A$ and $p_B$ related to $\phaseflipFunc$ through Eq.~\eqref{eq:phaseflipfunc2}.}
    \label{fig:Probability_distribution_Two_Ph-Flip}
\end{figure}

The fidelity probability density function when a perfect resource state is affected by local phase flip channels can then be obtained by substituting $\corrmat_\pf=\diag\{1-\phaseflipFunc,\phaseflipFunc-1,1\}$, Eq.~\eqref{eq:phaseflipfunc}, into Eq.~\eqref{eq:distrProbBellDiagonal}. Additionally, by calculating the eigenvalues of the resource state, we have that Bob's optimal operation is
$$\mathbb{w}_{l(\corrmat_d)}=\mathbb{w}_1,$$
for all value of $\phaseflipFunc$, therefore:
\begin{align*}
    a_1&=1, & a_2&=a_3=1-\phaseflipFunc.
\end{align*}
Thus, $F_{\min}^\pf =\frac{2-\phaseflipFunc}{2}$ is the minimal fidelity value that the protocol can produce, while the maximal is $F_{\max}^\pf=1$, noise independent. Consequently, the probability density function is:
\begin{align}\label{eq:distrProbBellDiagonal_phase_flip}
    \PDFBellDiagPhaseFlip(F)=\begin{cases}
        \frac{1}{\sqrt{\phaseflipFunc(2F-2+\phaseflipFunc)}}, & F\in\left(\frac{2-\phaseflipFunc}{2},1\right], \\
        0, & F\not\in(\frac{2-\phaseflipFunc}{2},1].
    \end{cases}
\end{align}
In Figure~\ref{fig:Probability_distribution_Two_Ph-Flip}, we plot this probability density function for different values of $\phaseflipFunc$. Another noteworthy feature is that local phase flip channels result in a symmetric distribution of fidelity probabilities for subsystems $A$ and $B$. This becomes clear because the probability density function is entirely determined by $\phaseflipFunc$, introduced in Eq.~\eqref{eq:phaseflipfunc2}, which is indeed symmetric.

This protocol produces a fidelity value within $[F_a,F_b]\subset[F_{\min}^\pf,1]$ with probability,
\begin{align}
    \label{eq:Prob_phase_flips}
    \ProbPhaseFlips(F\in[F_a,F_b])\!=\!\frac{\!\sqrt{2(F_b-1)\!+\!\phaseflipFunc}\!-\!\sqrt{2(F_a-1)\!+\!\phaseflipFunc}}{\sqrt{\phaseflipFunc}}.
\end{align}
The average fidelity is,
\begin{align}
    \label{eq:av_phase_flips}
    \avPhaseFlips\!=1-\frac{\phaseflipFunc}{3},
\end{align}
while the variance results in,
\begin{align}
    \label{eq:variance_phase_flips}
    \varPhaseFlips\!=\phaseflipFunc^2/45.
\end{align}
The previous average and variance values correspond to the probability density function in Eq.~\eqref{eq:distrProbBellDiagonal_phase_flip}.



A simpler scenario arises for local depolarizing channels. The resource state in this case is a Werner state with correlation matrix $\corrmat_\depol=\mw_1(1-\pa)(1-\pb)\mathbb{1}$, see Eq.~\eqref{eq:depol_resource_state_mat_corr}. As previously established in Ref.~\cite{Bussandri2024}, this resource state exhibits deterministic and symmetrical behavior. The fidelity probability density function, thus, collapses into a Dirac delta function:
\begin{align}\label{eq:probdistrDepol}
   \pdf^\depol(F) = \delta\!\left\{F-\left[\frac{2-\pa\pb}{2}\right]\right\}.
\end{align}

\subsubsection{Amplitude-damping noises}\label{sec:Amplitude_damping_noises}

Let us analyze the fidelity probability density function when a perfect Bell state ($\ket{\Phi_1}$) is affected by local amplitude-damping channels. The resource state is determined by Eqs.~\eqref{eq:TWOADCReducedStates} and \eqref{eq:TWOADCCOrrMat}. The Bloch vector of the output states $\outputStateJDataVect$, Eq.~\eqref{eq:outputState}, and their corresponding probabilities, Eq.~\eqref{eq:probSQT}, are, respectively,
\begin{align}
    \blochVectorOutputStates&=\frac{\pb[\mw_j]_{33}\hat k+\mw_1\corrmat_{\text{ad}}\vect}{1+\pa[\mw_j]_{33}t_3}, \label{eq:ADCBlochVectCondStates}\\
    \ProbJDadoVect&=\frac{1+\pa[\mw_j]_{33}t_3}{4}.
\end{align}
The fidelity can be calculated by the usual expression $F_j(\vec{t})=\frac{1}{2}(1+\vect\cdot\blochVectorOutputStates)$. 

As we can see from Eq.~\eqref{eq:ADCBlochVectCondStates}, the dependence of $F_j(\vec{t})$ with $j$ is ruled by $[\mw_j]_{33}$, which satisfies $[\mw_1]_{33}=[\mw_2]_{33}=1$ and $[\mw_3]_{33}=[\mw_4]_{33}=-1$. As $F_j(\vec{t})$ may assume only two different values, we re-indexed Alice's results by taking $[\mw_j]_{33} \to (-1)^k$, leading to $F_k(\vec{t})$ with $k\in\{0,1\}$.

The next result specifies the fidelity probability density function  derived in Appendix~\ref{app:TwoADCCalculation}.


\begin{figure*}
    \centering
    \includegraphics[width=0.8\linewidth]{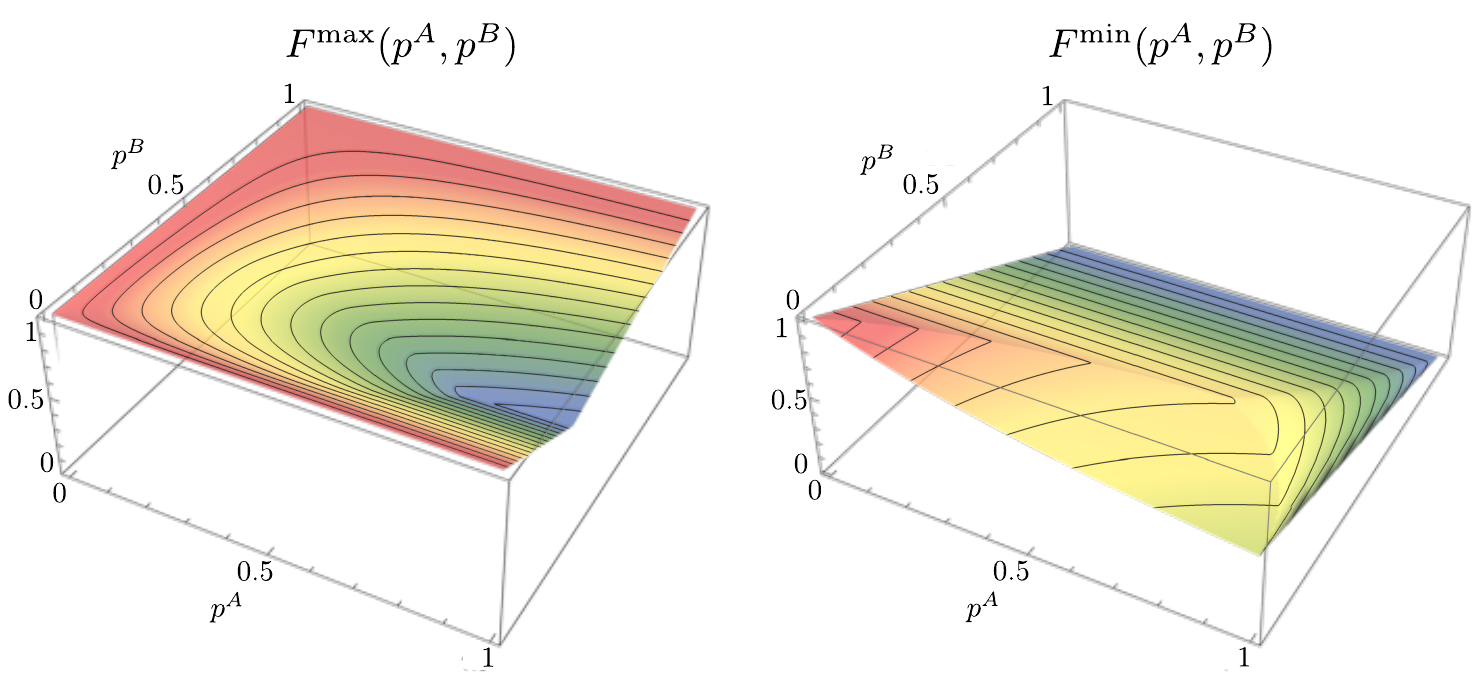}
    \caption{
Standard quantum teleportation protocol: Contour lines representing maximal $F^{\max}$ and minimal $F^{\min}$ fidelity values obtained for local amplitude-damping noise. These functions of noise parameters $p^A$ and $p^B$, do not depend on the outcome $k$ of the measurement, $F^{\min}(\pa,\pb) = F_{k}^{\min}(\pa,\pb)$ and $F^{\max}(\pa,\pb) = F_{k}^{\max}(\pa,\pb)$. Notably, these optimal values showcase the non-symmetry (with respect to $\pa\leftrightarrow\pb$) inherent to this noise model.
}
    \label{fig:ConditionalFidelityOpt}
\end{figure*}

\begin{result}\label{res:Distr}
    The probability density function of the fidelity for the standard quantum teleportation protocol with a Bell state $\ket{\Phi_1}$ affected by two local amplitude-damping channels, $\SQTADProtocol$, with noise parameters $\pa$ and $\pb$, is,
    \begin{align}\label{eq:distrprobtwoadc}
        \PDFADC(F)&=\sum_{k,l=0}^1 \PDFADC_{kl}(F), \\
    \PDFADC_{kl}(F)&=\begin{cases}
            \frac{1+(-1)^k z_{kl}(F)\pa}{4|\overlapADC'_k[z_{kl}(F)]|} & F\in \mathcal{F}_{k}^{\opt}(\pa,\pb) \nonumber \\
            0 & F\not\in\mathcal{F}_{k}^{\opt}(\pa,\pb), 
        \end{cases}
    \end{align}
    where $z_{kl}$ are such that $\overlapADC_k(z_{kl})=2F-1$, and,
    \begin{small}
        \begin{align}
        \overlapADC_k(z)\!&=\!\frac{(-1)^k\pb z\!+\!z^2 g(\pa,\pb) \!+\!\sqrt{\!(1\!-\!\pa)\!(1\!-\!\pb)\!}}{1+(-1)^k \pa z} \\
        g(\pa,\pb)&\!=\!1\!-\!\pa\!-\!\pb\!+\!2\pa\pb\!-\!\sqrt{(1-\pa)(1-\pb)} \nonumber \\
        \mathcal{F}_{k}^{\opt}(\pa,\pb)&=[F_{k}^{\min}(\pa,\pb),F_{k}^{\max}(\pa,\pb)] \\
        F_{k}^{\min}(\pa,\pb) &= \min_{\vect\in S} F_k(\vect), \ F_{k}^{\max}(\pa,\pb)= \max_{\vect\in S} F_k(\vect) \label{eq:fidelityMaxMinADC}
    \end{align}
    \end{small}

 Index $l$ labels solutions $z_{kl}\in[-1,1]$ such that $\overlapADC_k(z_{kl})=2F-1$. As this is, in principle, a quadratic equation, we take $l\in \{0,1\}$.
\end{result}


The minimal and maximal fidelity functions, denoted as $F^{\min}(\pa,\pb) = F_{k}^{\min}(\pa,\pb)$ and $F^{\max}(\pa,\pb) = F_{k}^{\max}(\pa,\pb)$, see Eq.~\eqref{eq:fidelityMaxMinADC}, respectively, are independent of Alice's measurement result, $k$. These functions are visualized in Fig.~\ref{fig:ConditionalFidelityOpt}. 

\begin{figure*}[htb]
    \centering
    \includegraphics[width=\textwidth]{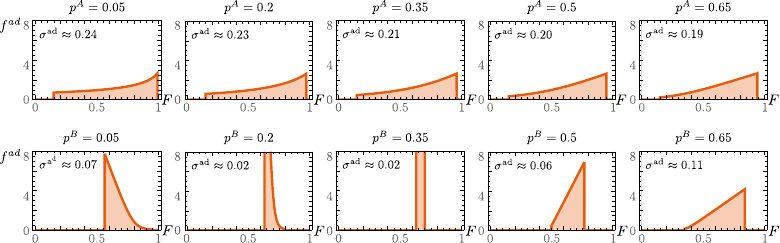}
    \caption{Fidelity probability density function, $\PDFADC(F)$, Eq.~\eqref{eq:distrprobtwoadc}, for different local amplitude-damping noises $\pa$ and $\pb$: In the upper (lower) figures, noise is fixed in the system $B$ ($A$), $\pb=0.85$ ($\pa=0.85$), and increased in system $A$ ($B$). Plots vertically aligned (i.e., in the same column) display probability density functions that imply equal average fidelities despite having highly different statistical behaviors. $\sigma^{\text{ad}}$ is the fidelity standard deviation in each corresponding case.}
    \label{fig:probdristrTWOADC}
\end{figure*}

Previous studies have characterized the case of teleportation under amplitude-damping noises as exhibiting \textit{symmetric} fidelity behavior, relying on average fidelity values~\cite{Knoll2014,Fortes2015}. However, a closer examination of the probability density functions reveals a highly non-symmetric fidelity behavior. Figure~\ref{fig:probdristrTWOADC} showcases this non-symmetry, illustrating the evolution of the probability density function (Eq.~\eqref{eq:distrprobtwoadc}) as noise is applied to one subsystem but with fixed noise ($p^*=0.85$) in the other. Specifically, the upper (lower) figures show how the fidelity probability density function changes as noise is applied in system $B$ ($A$) while in system $A$ ($B$) remains fixed. Remarkably, those upper and lower distributions corresponding to the same column have equal average fidelity but highly different statistical behavior. Lastly, for $\pa=\pb=1$, the resulting probability density function converges to that of the optimal classical protocol: $2F$ (Eq.~\eqref{eq:probDistrClassProtOpt}).

\section{Quality assessment of teleportation}
\label{sec:prior_importance_certification}

\begin{figure}[htb]
   \centering
   \includegraphics[width=.8\linewidth]{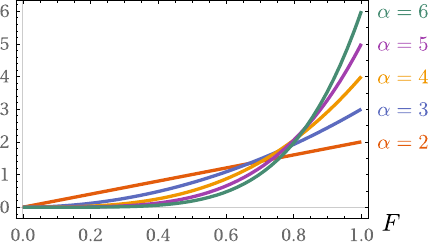}
   \caption{The importance prior function $\importancePriorBeta{\alpha}{1}(F)$, Eq.~\eqref{eq:importanceprior}, for different values of $\alpha$ and $\beta=1$. Higher values of $\alpha$ assign greater importance to higher fidelity values.}
   \label{fig:importanceprior}
\end{figure}

\begin{figure*}[htb]
    \centering
    \includegraphics[width=\textwidth]{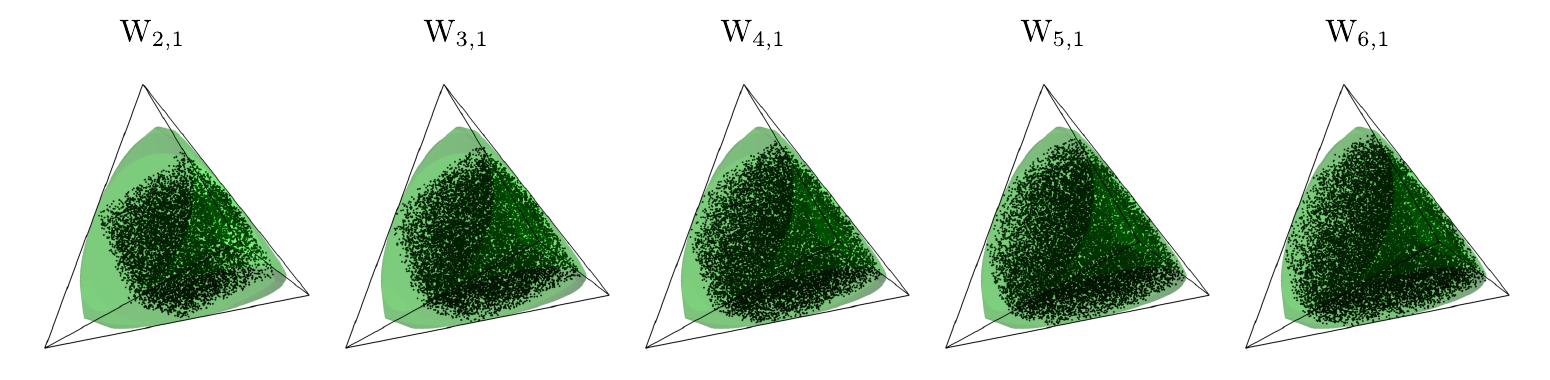}
    \caption{Tetrahedron of Bell-diagonal states: Set of diagonal correlation matrices $\corrmat$ [Eq.~\eqref{eq:BellDiagonalStates}] corresponding to physical states. The green region stands for Bell-diagonal states satisfying the CHSH inequality. Black dots represent Bell-diagonal states leading to a non-successful certification ($\gamma_{\importancePriorBeta{\alpha}{\beta}}(\PDFBellDiag,\PDFClassicalProtOpt)<0$, see Eq.~\eqref{eq:criteriaPriorImportance}), according to different importance prior functions $\{\importancePriorBeta{\alpha}{1}\}_{\alpha=1}^6$, Eq.~\eqref{eq:importanceprior}. As the parameter $\alpha$ is increased, the teleportation assessment becomes more rigorous by assigning greater priotiry to higher fidelity values, thereby improving the teleportation quality (see Fig.~\ref{fig:importanceprior}). We generated $20000$ points, uniformly distributed in the tetrahedron, and black colored only if they were non-successful. We see that as we assign more importance to higher fidelity values, more entangled states lead to a non-successful protocol. Non-local states may seem to hold a successful certification even for $\importancePriorBeta{6}{1}$.}
    \label{fig:bdUseful}
\end{figure*}

\begin{figure}[htb]
    \centering
    \includegraphics[width=0.4\textwidth]{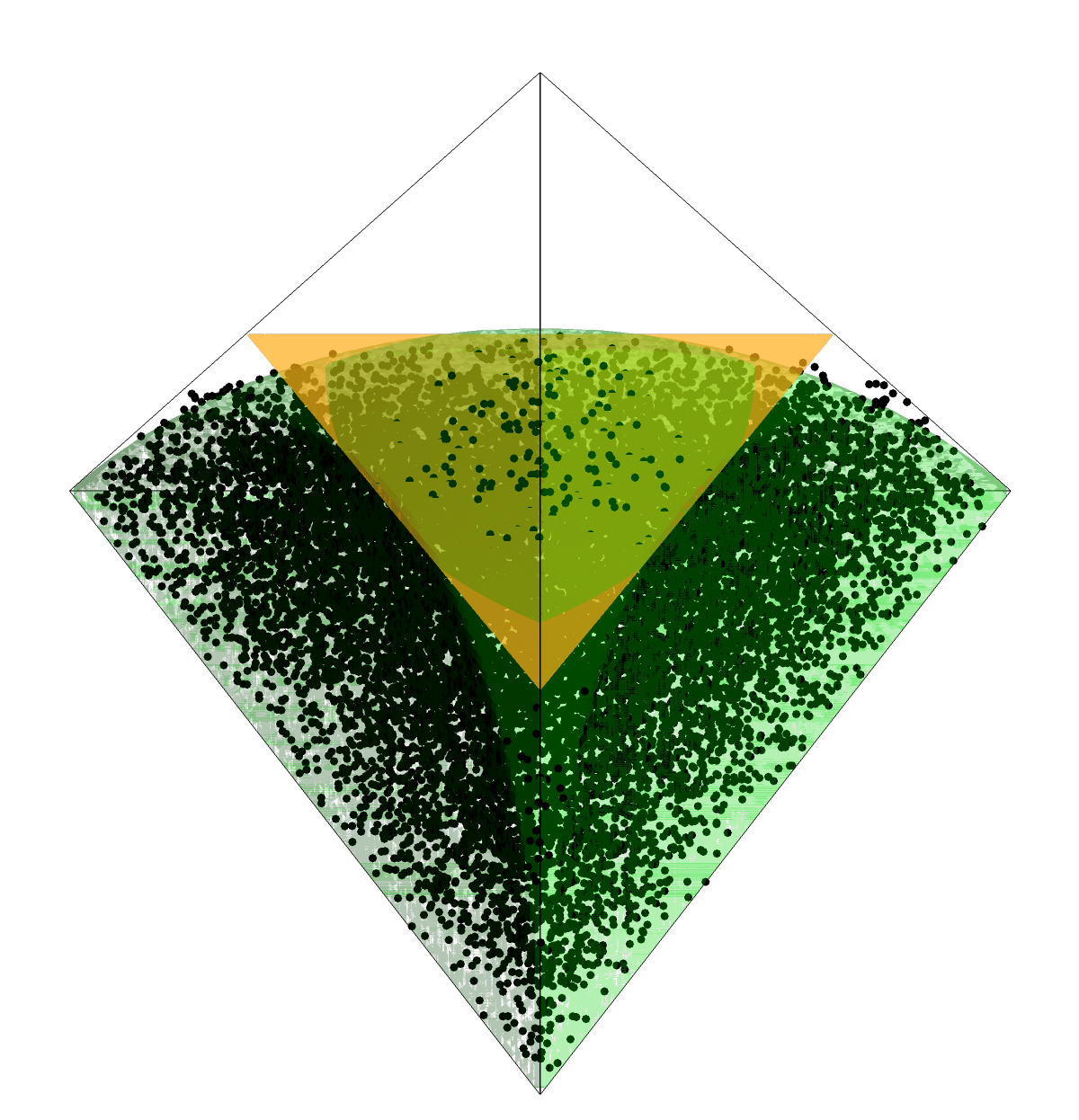}
    \caption{
    Amplification of one vertex of the tetrahedron spanned by Bell states (see Figure~\ref{fig:bdUseful}). The green region represents the set of states satisfying the CHSH inequality. The yellow surface indicates a constant average fidelity of $0.8$. All depicted quantities are dimensionless.
    We generated $20000$ points, uniformly distributed in this region, and colored as black those leading to non-successful certification according to the prior $\importancePriorBeta{8}{1}$, ($\gamma_{\importancePriorBeta{8}{1}}(\PDFBellDiag,\PDFClassicalProtOpt)<0$, see Eqs.~\eqref{eq:criteriaPriorImportance} and~\eqref{eq:importanceprior}).
    }
    \label{fig:plot_BD_W8}
\end{figure}

Through the analysis in Section~\ref{sec:Teleportation_fidelity_statistics}, we find that standard quantum teleportation protocols encompass a wide spectrum of fidelity statistical behaviors, extending from deterministic to highly variable outcomes. The most employed criterion for successful teleportation, as discussed in Sec.~\ref{sec:Teleportation_protocols}, relies on the first moment of the fidelity distributions, specifically, exceeding the measurement-prepare average fidelity, Eq.~\eqref{eq_classical_av_fid}. However, this criterion can be overly simplistic, potentially masking contentious nuances in protocol performance. For example, consider the fidelity probability density function of the optimal classical teleportation protocol, Eq.~\eqref{eq:probDistrClassProtOpt}, and the one arising from standard quantum teleportation affected by local depolarizing noises, Eq.~\eqref{eq:probdistrDepol}.  The classical protocol consistently yields an almost flat distribution, indicating a broad range of possible fidelity values and a high standard deviation.  In contrast, the quantum protocol always produces the fidelity value determined by the noise level $(1-\pa)(1-\pb)$. When this parameter is negligibly greater than $\braket{F}^\cl_\opt=2/3$, the protocol is classified by the average-based criterion as \textit{successful}, despite never achieving a fidelity higher than, for example, $0.7$. Conversely, the classical protocol has a $0.51$ probability of attaining a fidelity $F > 0.7$, with the most probable value being $F=1$ (perfect teleportation). On the other hand, if we focus exclusively on obtaining high fidelity values, we would motivate the adoption of an alternative criterion based on the probability, for example, $\text{Pr}\{F\in[0.7,1]\}$, which would deem the noisy depolarized quantum protocol \textit{unsuccessful}.

Another peculiar effect arises for amplitude-damping local noises~\cite{Knoll2014,Fortes2015}: Fig.~\ref{fig:probdristrTWOADC} shows distinct fidelity probability density functions that, while maintaining identical average fidelity values within each column, exhibit markedly different statistical characteristics. The distributions range from nearly uniform probability profiles (upper panels) to highly non-uniform cases characterized by substantial regions of zero probability density (lower panels). This case implies a highly non-symmetric behavior, which is not exhibited by the average fidelity alone.  Additionally,  as illustrated by the left subfigure in Fig.~\ref{fig:contourplotADC}, increasing noise in one subsystem can increase the average fidelity within certain noise regimes, leading to a successful certification by the average-based criterion.


By having the full fidelity probability density function, as calculated in Sec.~\ref{sec:Teleportation_fidelity_statistics}, we can introduce deeper criteria for \textit{high-fidelity} successful teleportation based on \textit{prior importance} functions.

A prior importance function assigns a weight to each possible fidelity value, prioritizing higher fidelity outcomes over others. Together with the corresponding probability density functions, prior importance functions allow us to rewrite all successful teleportation criteria in terms of a weighted performance difference between an arbitrary protocol $P$ and the classical one $\classicalProt$. 

\begin{definition} \label{def:newcriterion}
Let $W$ be a continuous probability distribution in $[0,1]$. A protocol $P$ is \textit{successful according to $\importancePrior$} if and only if $\gamma_\importancePrior(P,\classicalProtOpt)>0$, with
\begin{align}\label{eq:criteriaPriorImportance}
   \gamma_\importancePrior(P,\classicalProtOpt)=\int_0^1 df \importancePrior(F) [\pdf(F)-\PDFClassicalProtOpt(F)].
\end{align}
where $\classicalProtOpt$ is the optimal measurement-prepare protocol with probability density function $\PDFClassicalProtOpt(F)=2F$, see Eq.~\eqref{eq:probDistrClassProtOpt}.
\end{definition}
This approach provides a general way to assess performance by accounting for the relative importance of different fidelity outcomes. For example, when the prior importance function $W$ corresponds to the average fidelity calculation, i.e., $W=W_\cl=2F$, we recover the usual average-based criterion. Conversely, if $W$ is a piecewise step function, we obtain a probability-based criterion that depends on the cumulative distribution function $\text{Pr}\{F\in [F_t,1] \}$, where $F_t$ represents a threshold that must be overcome.

A natural way to generalize the prior importance function is to use Beta distributions:
\begin{equation}\label{eq:importanceprior}
   \importancePriorBeta{\alpha}{\beta}(F)=\frac{F^{\alpha-1}(1-F)^{\beta-1}}{\text{B}(\alpha,\beta)},
\end{equation}
where $\text{B}(\alpha,\beta)$ is the Beta function. When $\alpha>1$, $\importancePriorBeta{\alpha}{\beta}(0)=0$. Conversely, when $0<\alpha<1$, the function approaches infinity as $F$ approaches zero.  If $\beta>1$, the function is zero at $F=1$, while if $0<\beta<1$, it approaches infinity as $F$ approaches $1$.  The special case where $\beta=1$ and $\alpha>1$ is particularly interesting for us, as it assigns more importance to higher fidelity values with $\alpha$. Fig.~\ref{fig:importanceprior} showcases this behavior. Additionally, for $\beta=1$ and $\alpha=2$, the function simplifies to $\importancePriorBeta{2}{1}(F) = 2F$, equivalent to the usual average-based criterion.  Furthermore, $\importancePriorBeta{3}{1}(F)=3F^2$ can be interpreted as a deviation-based criterion because it relates to the second moment of the probability density function.

\subsection{Bell-diagonal states}
\label{sec:certification_Bell_diagonal}

We now examine how different prior importance functions, $\importancePriorBeta{\alpha}{1}$, Eq.~\eqref{eq:importanceprior}, certify noisy quantum teleportation protocols with a Bell-diagonal resource state, see Sec.~\ref{sec:standard}, Eq.~\eqref{eq:BellDiagonalStates}, according to the new criterion in Def.~\ref{def:newcriterion}, Eq.~\eqref{eq:criteriaPriorImportance}. To do this, we generated random states of this kind and computed the criterion defined by $\gamma_{\importancePriorBeta{\alpha}{1}}$, see Eq.\eqref{eq:criteriaPriorImportance}. 

In Figure~\ref{fig:bdUseful} we showcase non-successful states (black dots) inside the tetrahedron of Bell-diagonal states (the three-dimensional representation whose four vertices correspond to the maximally entangled Bell states).
%
%
For $\alpha=2$, the usual average-based criterion is obtained, and we can see that, as it is already known, the non-successful states fall inside the octahedron of separable states. However, as we increase the teleportation quality by assigning more importance to higher fidelity values, more entangled states become non-successful. A remarkably robust quantum state category is non-local two-qubit quantum states, defined by violating the Clauser–Horne–Shimony–Holt inequality (green region)~\cite{Clauser1969}, as they seem to remain \textit{successful} even according to $\importancePriorBeta{6}{1}$. 

In Figure~\ref{fig:plot_BD_W8} we evaluate our new criteria $\gamma_{\importancePriorBeta{8}{1}}$ for entangled Bell-diagonal states (we showcase one of the corresponding tetrahedrons because by symmetry the other ones are identical), showing additionally the contour plot of the average fidelity $\braket{F}^{\text{Bd}}=0.8$, and the green region of states satisfying the CHSH inequality. This figure suggests that within the same surface of constant average fidelity, non-local states exhibit high-fidelity successful teleportation (even according to $\importancePriorBeta{8}{1}$) while other states do not. This observation deserves further exploration, which we will address in future work.


\subsection{Amplitude-damping noise} \label{sec:Amplitude_damping_noises_analysis}
\label{sec:certification_ADC}

\begin{figure*}[htb]
    \centering
    \includegraphics[width=\textwidth]{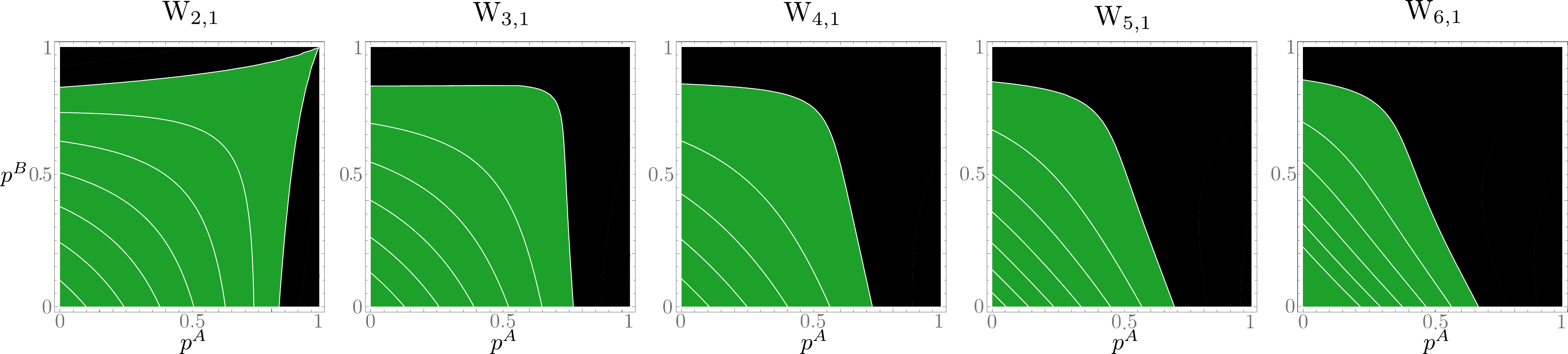}
    \caption{
    Contour plots of the function $\gamma_{\importancePriorBeta{\alpha}{\beta}}(\PDFADC,\classicalProt)$ [Eq.\eqref{eq:criteriaPriorImportance}], which defines our new criterion for successful teleportation (see Definition\ref{def:newcriterion}). The protocol under analysis is the standard quantum teleportation scheme using the Bell state $\ket{\Phi_1}$ as a resource, subject to local amplitude-damping noise with damping parameters $(\pa,\pb)$. The associated fidelity probability density function is $\PDFADC$ [Eq.~\eqref{eq:distrprobtwoadc}]. Green (black) regions correspond to certified (non-certified) teleportation. For the prior importance function $\importancePriorBeta{2}{1}$ (left plot), the contour lines coincide with those obtained from the average fidelity, as reported for example in Ref.~\cite{Fortes2015}. This reproduces the \textit{fighting noise with noise} effect, where increasing the noise in one subsystem can improve certification. However, we see that for higher-quality criteria ($\alpha > 2$), this effect disappears and the green regions shrink. Additionally, increasing noise in subsystem $A$ is shown to be more detrimental than in subsystem $B$.
    }
    \label{fig:contourplotADC}
\end{figure*}

Let us focus on verifying noisy quantum teleportation under local amplitude-damping channels, see Sec.~\ref{sec:Amplitude_damping_noises}, according to our new criteria defined by $\gamma_{\importancePriorBeta{\alpha}{1}}$, Eq.~\eqref{eq:criteriaPriorImportance}. The probability density function to be analyzed in this case is $\PDFADC$, Eq.\eqref{eq:distrprobtwoadc}. 

Figure~\ref{fig:contourplotADC} displays the noise parameter region $(\pa,\pb)$ for which the protocol is successful (green) for $\alpha\in\{2,\dots,6\}$ and $\beta=1$. Black regions indicate that the noisy protocol is not successful. 

The first plot on the left side corresponds to $\alpha=2$, i.e. the usual average-based criterion. Two facts are noteworthy in this plot: 1) There exist intervals such that increasing noise in one subsystem leads from the black to the green region, an effect referred to as \textit{fighting noise with noise}~\cite{Badziag2000,Bandyopadhyay2000,Knoll2014,Fortes2015}; 2) The average-based criterion implies a fully symmetrical certification in the sense that would be the same increasing the noise in one subsystem or the other. 

However, these two features are exclusively attached to the average-based criterion ($\alpha=2$ and $\beta=1$): In Sec.~\ref{sec:Amplitude_damping_noises}, we computed the fidelity probability density function, showing that fidelity behaves in a highly non-symmetrical way. Alternatively, Fig.~\ref{fig:contourplotADC} showcases that whenever we assign more importance to higher fidelity values by increasing $\alpha>2$, more noise in one subsystem does not lead to a \textit{successful} certification in any case. These results suggest that the so-called 'fighting noise with noise' effect is merely an artifact introduced by the choice of prior importance function, rather than a genuine advantage. Additionally, standard quantum teleportation does not behave symmetrically under local amplitude-damping noises, and its performance is most robust under increasing noise in $B$ than in subsystem $A$.

\section{Concluding remarks}
\label{sec:conclusions}

Every proposed device or protocol intended to teleport quantum states must be subjected to an assessment or evaluation phase, probing its capabilities. This work aims to go further into the quality assessment of single-qubit teleportation protocols.

Given a use of the protocol to be evaluated, the fidelity between the input and output states stands as the key element in this analysis. This defines a random variable containing the necessary information to assess the protocol's quality. 

The average fidelity coincides with the mean fidelity value over infinite uses of the protocol. Because, in some cases, its calculation for pure input states is simple and it can be experimentally estimated by employing just the six states comprising the mutually unbiased basis in dimension two (or tomographically complete basis), the average fidelity is still nowadays the most employed figure of merit to certify successful teleportation~\cite{valivarthi_quantum_2016,ren_ground-to-satellite_2017,valivarthi_teleportation_2020,zuo_overcoming_2021,ribeiro_finite_2024,liu_overcoming_2024,sala_quantum_2024,ribeiro_detecting_2024,polacchi_teleportation_2024,gong_optimal_2024,masajada_optimizing_2025,wu_does_2023,xu_noise_2024}. 
However, as we pointed out, this single-number metric may mask drastically different statistical behaviors: For experimentalists seeking to reliably produce high-fidelity outcomes, certain protocols may prove more advantageous than others, even when exhibiting identical average fidelities.


In light of these limitations, we developed a comprehensive statistical framework to assess quantum teleportation performance beyond the conventional average-fidelity benchmark. Our main contributions are organized into two parts. The first one, as described in Sec.~\ref{sec:Teleportation_fidelity_statistics}, represents the full statistical characterization of fidelity in teleportation protocols. In Result~\ref{res:fidelity_prob_distr}, we provided an analytical expression of the fidelity probability density function in terms of the ensemble of conditional states defining the protocol. Result~\ref{res:probDistrClassProt} establishes the probability density function for the classical, or measurement-and-prepare, protocol. As will become clear later, this is a crucial element for our proposed quality assessment of teleportation. The next protocol to analyze is standard quantum teleportation for two different resource states: a) Bell diagonal states, whose fidelity probability density function is given by the complete elliptic integral
of the first kind $K$, see Result~\ref{res:BD_prob_distr}; and b) a perfect Bell state affected by two local amplitude-damping noises (Result~\ref{res:Distr}).

Building on this statistical foundation, the second part is devoted to the quality assessment of teleportation protocols. 
The main idea is to employ the fidelity probability density function to define more meaningful and deeper criteria for successful teleportation. In this line of reasoning, we proposed in Sec.~\ref{sec:prior_importance_certification} to use prior importance functions to prioritize certain fidelity ranges, leading to a modular and generalized successful teleportation criterion, Def.~\ref{def:newcriterion}. A natural way to assess high-fidelity teleportation is through the use of beta distributions, Eq.~\eqref{eq:importanceprior}. In the next two sections,~\ref{sec:certification_Bell_diagonal} and~\ref{sec:Amplitude_damping_noises_analysis}, we apply our new criteria to assessing standard quantum teleportation with Bell diagonal resource states, and to the case of perfect Bell states affected by local amplitude-damping noises.

For Bell-diagonal sources, we presented a numerical exploration of the capabilities of this kind of states, increasing the teleportation quality by prioritizing higher fidelity values (i.e., increasing $\alpha$, see Fig.~\ref{fig:bdUseful}). This numerical analysis suggests that non-local states, i.e. those not satisfying the CHSH inequality, represent a remarkably robust set of resource states in the sense that they seem to assure a successful high-fidelity teleportation, even for $\alpha=6$. 


Our framework also clarifies the ``fighting noise with noise'' phenomenon reported for amplitude-damping channels~\cite{Badziag2000,Bandyopadhyay2000,Knoll2014,Fortes2015}. Figure~\ref{fig:probdristrTWOADC} reveals that this effect stems from highly asymmetric fidelity distributions that, despite their different shapes, share the same average fidelity. While the standard average-based criterion ($\alpha=2$) may certify these scenarios as successful, our analysis shows this is an artifact of the chosen metric. As illustrated in Fig.~\ref{fig:contourplotADC}, for any stricter criterion that prioritizes high-fidelity outcomes ($\alpha>2$), this apparent advantage vanishes, and the protocol is no longer certified as successful. This approach uncovers a hidden asymmetry in the protocol's performance, demonstrating that it is more robust against noise in subsystem $B$ than in subsystem $A$.

Finally, although these new and more stringent criteria for high-fidelity teleportation are crucial for developing better teleportation protocols, it is necessary to explore how they can be experimentally estimated. The six states comprising the mutually unbiased basis in the qubit case, usually employed to estimate the average fidelity, are not sufficient for our new criteria, and the full Haar measure, in principle, has to be emulated. An interesting starting point would be the case of the importance prior $W_{3,1}$, which leads to a criterion that can be estimated if the fidelity variance (the actual variance, not the channel-fidelity deviation as calculated in Ref.~\cite{Ghosal2020a}) is experimentally estimated. 

\begin{acknowledgments}
The authors gratefully acknowledge Luis Miguel Nieto Calzada for his valuable contributions to this work.
DGB was supported by the Q-CAYLE project, funded by the
European Union-Next Generation UE/MCIU/Plan de Recuperación, Transformación y Resiliencia/Junta de Castilla y Leon (PRTRC17.11), and also by RED2022134301-T and PID2023-148409NB-I00, financed by MICIU/AEI/10.13039/501100011033. The financial support of the Department of Education of the Junta de Castilla y León and FEDER Funds is also gratefully acknowledged (Reference: CLU-2023-1-05). GMB acknowledges financial support from project PIBAA 0718 funded by Consejo Nacional de Investigaciones Cient\'ificas y T\'ecnicas CONICET (Argentina).
\end{acknowledgments}

\newpage
\onecolumngrid
\appendix

\section{Fidelity distribution}\label{app:Distr}

In probability theory, the characteristic function of an arbitrary random variable $X$ is $G(\kappa)=\braket{\exp i\kappa x}$, being $x$ an outcome of $X$. In our case, the characteristic function of the fidelity $F_j(\vec{t})$, given in Eq.~\eqref{eq:PosteriorFidelityCont}, reads,
\begin{align}
    G_F(\kappa)=\braket{\exp i\kappa F}=\int_S \sum_j \frac{1}{4\pi}d\Omega \ProbJDadoVect \exp i\kappa F_j(\vec{t} \, ). \label{eq:characFuncPriorFid}
\end{align}

We can calculate the one-parameter probability density function of obtaining the fidelity value $F_{j}(\vect)=F$ from their characteristic functions by taking their Fourier transform:
\begin{align}
   \pdf(F) = \frac{1}{2\pi} \int_{-\infty}^{\infty} G_{F}(\kappa) e^{-i\kappa F} d\kappa. \label{eq:probabilityDistrPosteriorFid}
\end{align}
By inserting Eq.~\eqref{eq:characFuncPriorFid} into the corresponding previous equations and using the   identity $\delta(x)=\frac{1}{2\pi}\int_{-\infty}^{\infty} d\kappa \exp i\kappa x$, we arrive at:
\begin{align}
   \pdf(F) &= \sum_j \pdf_j(F) \nonumber \\
   \pdf_j(F) &=\frac{1}{4\pi}\int_S d\Omega \ \ProbJDadoVect \ \delta\!\left[F_j(\vec{t})-F\right] \label{eq:probabilityDistrDelta}
\end{align}
Now, let us define the sets $S_j=\{\vec{x}\in S; F_j(\vec{x})=F\}$, with $S$ the Bloch sphere. By assuming that the function $F_j$ is $C^{\infty}(S)$, the composition of the Dirac delta distribution with the smooth map $F_j(\vec{t})-F$ implies (see Theorem 6.1.5 in Chapter IV of Ref.~\cite{Hormander1983}),
\begin{align} \label{eq:ProbDistrEnGeneral}
   \pdf_j(F) =\frac{1}{4\pi}\int dS_j \  \frac{p_{j|\vec{x}}}{\left|\nabla F_j(\vec{x})\right|},
\end{align}
where $dS_j$ stands for the Euclidean surface measure on the surface $S_j$. 

The previous statement does not fix any particular parametrization. A natural choice is to use spherical coordinates: $t_1=\cos\phi\sin\theta$,  $t_2=\sin\phi\sin\theta$, and $t_3=\cos\theta$. The corresponding metric elements are $g_\phi=(\sin\theta)^2$, and $g_\theta=1$. 

The set $S_j$ is defined by those vectors $\vec{x}=\vec{x}(\phi,\theta)$ satisfying the surface equation $F_j(\vec{x})=F$. The gradient in spherical coordinates is $\nabla F_j(\vec{x})=\hat{e}_\phi \frac{\partial_\phi F_j(\vec{x})}{\sin\theta}+\hat{e}_\theta \partial_\theta F_j(\vec x)$, where $\{\hat e_\phi, \hat e_\theta\}$ defines the usual normalized covariant basis with constant radial coordinate.

Let us suppose that $\theta$ is constant and we solve for $\phi$, then: 1) Let $\phi_i(\theta,F)$ denote the roots, i.e. $F_j[\vec x_i(\theta,F)]-F=0$, where $\vec x_i(\theta,F)=\vec x[\phi_i(\theta,F),\theta]$; 2) $\left|\nabla F_j(\vec{x})\right|=\left|\frac{\partial_\phi F_j(\vec{x})}{\sin\theta}\right|$; and 3) the differential surface element $dS_j$ is, $dS_j=\sqrt{g_\theta}d\theta=d\theta$. Thus, Eq.~\eqref{eq:ProbDistrEnGeneral} becomes: 
\begin{align} \label{eq:ProbDistrEnGeneralPhi}
   \pdf_j(F) =\frac{1}{4\pi}\int_{0}^\pi d\theta \sin\theta \ \sum_i \frac{p_{j|\vec{x}_i}}{\left|\partial_\phi F_j(\vec{x}_i)\right|}.
\end{align}

If, on the other hand, we solve for $\theta$ assuming constant $\phi$, by following the same reasoning as before, we arrive at that Eq.~\eqref{eq:ProbDistrEnGeneral} results in:
\begin{align} \label{eq:ProbDistrEnGeneralTheta}
   \pdf_j(F) =\frac{1}{4\pi}\int_{0}^{2\pi} d\phi \sum_i \sin \theta_i(\phi,F) \  \frac{p_{j|\vec{x}_i}}{\left|\partial_\theta F_j(\vec{x}_i)\right|},
\end{align}
where, in this case, $\vec x_i=\vec x_i(\phi,F)=\vec x [\phi,\theta_i(\phi,F)]$ satisfying $F_j[\vec x_i(\phi,F)]-F=0$.

\section{Fidelity distribution for classical protocols} \label{app:ClassicalDistr}
In Sec.~\ref{sec:measurementPrepare}, we introduced arbitrary classical protocols. They are characterized by the ensemble $\{\ProbJDadoVect^\cl,\rho^{\cl,B}_{j|\vec{t}}\}$,
\begin{align*}
    \rho^{\cl,B}_{j|\vec{t}} &= \frac{1}{2} \left(\Id + \vecr_j \cdot \vecsigma \right),\\
    \ProbJDadoVect^\cl &= \frac{c_j^2}{2} \left(1 + \vect \cdot \vecs_j\right),
\end{align*}
see Eqs.~\eqref{eq:ClassicalOutputs} and~\eqref{eq:probabilidades_clasicas}, respectively, and Eqs.~\eqref{eq:POVM_def} and~\eqref{eq:POVM_conditions} for the POVM conditions over the elements $\{c_i^2,\vec s_i\}_i$. Thus, the fidelity (Eq.~\eqref{eq:PosteriorFidelityCont}) is,
\begin{align}
    F_{j}(\vect)=\frac{1}{2}(1+\vect\cdot\vecr_j).
\end{align}
The first observation is that $\frac{1-r_j}{2}\leq F_j(\vec{t})\leq \frac{1+r_j}{2}$ for all $\vect\in S$, so the probability density function will be null outside of this range.

Now, let $\mathbb{O}^\cl$ be the rotation such that $\mathbb{O}^\cl\vecr_j=r_j\hat k$, it follows,
$$F_{j}(\vect)=\frac{1}{2}[1+r_j(\mathbb{O}^{\cl})^\intercal\vect\cdot\hat k].$$
Because of the spherical symmetry involved in the calculation of the probability density function, see for example Eq.~\eqref{eq:probabilityDistrDelta}, we can neglect the rotation $\mathbb{O}^\cl$ and take, simply,
\begin{align}
    F_{j}(\vect)=\frac{1}{2}(1+r_j\vect\cdot\hat k)=\frac{1}{2}(1+r_jt_3).
\end{align}
Let us solve directly for $\theta$,
\begin{align}
    F_{j}(\vect)=F \ \iff \ \theta(F) =\arccos \frac{2F-1}{r_j}.
\end{align}
Then, the probability density function is given by Eq.~\eqref{eq:ProbDistrEnGeneralTheta}. As,  
\begin{align}
    |\partial_\theta F_j(\vec{x})|=\frac{r_j}{2}\sin\theta,
\end{align}
it follows,
\begin{align}
  \frac{1}{4\pi}\int_{0}^{2\pi} d\phi \sum_i \sin \theta_i(\phi,F) \  \frac{p_{j|\vec{x}_i}}{\left|\partial_\theta F_j(\vec{x}_i)\right|} &=\frac{1}{2\pi}\int_{0}^{2\pi} d\phi \frac{p_{j|\vec{x}_i}}{r_j}=\frac{1}{2\pi}\int_{0}^{2\pi} d\phi \frac{c_j^2 \left(1 + \vect \cdot \vecs_j\right)}{2r_j}\\
  &=\frac{1}{2\pi}\int_{0}^{2\pi} d\phi \frac{c_j^2 \left[1 + s_{j,1}\cos\phi\sin \theta(F)+s_{j,2}\sin\phi\sin \theta(F)+s_{j,3}\cos\theta(F)\right]}{2r_j} \\
  &=\frac{c_j^2 \left[1+s_{j,3}\cos\theta(F)\right]}{2r_j}=\frac{c_j^2}{2r_j}\!\left[\!1\!+\!(2F-1)\!\frac{\vec{r}_j\cdot\vec{s}_j}{r_j ^2}\right].
\end{align}
Therefore, we arrive at, 
\begin{align}
    \PDFClassicalProt(F)&=\sum_{j} \PDFClassicalProt_{j}(F), \ \text{ where }   \pdf_{j}^{\cl}(F)=
        \begin{cases}
         \!\frac{c_j^2}{2r_j}\!\left[\!1\!+\!(2F-1)\!\frac{\vec{r}_j\cdot\vec{s}_j}{r_j ^2}\right], & \!F\in[\frac{1-r_j}{2},\frac{1+r_j}{2}] \\
        0, & \!F\not\in [\frac{1-r_j}{2},\frac{1+r_j}{2}].
        \end{cases}
\nonumber
\end{align}

\section{Quadratic form distribution}\label{app:QuatraticFormDistr}

Let us consider a random variable
\begin{align}\label{eq:appQuatraticForm}
    X=\frac{1}{2}(1+\vec{t}\cdot \mathbb{A}\vec{t}),
\end{align}
being $\vec{t}\in S$ uniformly distributed in the Bloch sphere, and $\mathbb{A}$ being a diagonalizable real matrix, $\mathbb{A}=\mathbb{O} \mathbb{A}_d \mathbb{O}^\intercal$, $\mathbb{A}_d=\text{diag}\{a_1,a_2,a_3\}$, with eigenvalues $|a_l|\in (0,1]$, $l=1,2,3$.

Let us assume the order $a_1 \! >\! a_2\!>\!a_3$. Then, it holds $a_1 \geq X \geq a_3$ for all $\vec{t}\in S$. 

Let $x$ be a fixed value of $X$ such that $2x-1\not=a_2$. Eq. \eqref{eq:appQuatraticForm} implies
$$x=\frac{1}{2}(1+\vec{t}_o\cdot \mathbb{A}_d\vec{t}_o),$$
with $\vec{t}_o = \mathbb{O} \vec{t}$. By the usual angle parametrization, $\vec{t}_o=\cos\phi\sin \theta  \hat{i}+ \sin \phi \sin \theta \hat{j}+ \cos \theta \hat{k}$, $\theta\in [0,\pi]$, and $\phi\in [0,2\pi)$. 

Now, let us eliminate $\phi$: We have four solutions $\phi_{ij}(x)=(-1)^j\arccos (-1)^i g(x,\mathbb{A}_d)$, $i,j\in \{0,1\}$, with, 
\begin{align}
    g(x,\mathbb{A}_d) &= \sqrt{\frac{2x-1-a_2+(a_2-a_3)(\cos\theta)^2}{(a_1-a_2)(\sin\theta)^2}}, \text{ if } s^2_{\min}(x,\mathbb{A}_d)\leq (\cos \theta) ^2 \leq s^2_{\max}(x,\mathbb{A}_d), \\
    s_{\min}^2(x,\mathbb{A}_d)&=\frac{a_2-(2x-1)}{a_2-a_3}, \\
    s_{\max}^2(x,\mathbb{A}_d)&=\frac{a_1-(2x-1)}{a_1-a_3}.
\end{align}
Additionally, 
\begin{align*}
    \left| \left.\partial_\phi X \right|_{\phi=\phi_{ij}} \right| = \sqrt{\left[(a_2-a_3)(\cos\theta)^2-(a_2-2x+1)\right]\left[a1-2x+1-(a_1-a_3)(\cos\theta)^2\right]}=\left|X'(x,\mathbb{A}_d,\cos\theta)\right|.
\end{align*}

Following Sec. \ref{app:Distr}, the probability density function is given by \eqref{eq:probabilityDistrDelta} and Eq. \eqref{eq:ProbDistrEnGeneralPhi}, in the deterministic case, i.e. when the random variable depends only on $\vec{t}$ distributed uniformly in the unit sphere. Taking $s=\cos \theta$, the probability density function is, therefore,
\begin{align}
    \pdf(x)&=\frac{1}{4\pi}\int_0^{\pi} d\theta \sin \theta \sum_{i,j=0}^1 \frac{1}{\left| \left.\partial_\phi X \right|_{\phi=\phi_{ij}(x)} \right|} = \frac{1}{\pi}\int_{s_{\min}(x,\mathbb{A}_d)}^{s_{\max}(x,\mathbb{A}_d)} ds \left|X'(x,\mathbb{A}_d,s)\right|^{-1}=\\
    &=\begin{cases}
        \frac{2}{\pi\sqrt{(a_2-a_3)(1+a_1-2x)}}K\left[\frac{(a_1-a_2)(2x-1-a_3)}{(a_2-a_3)(a_1-2x+1)}\right], & x\in\left[\frac{1+a_3}{2},\frac{1+a_2}{2}\right], \\
        \frac{2}{\pi\sqrt{(a_1-a_3)(2x-1-a_2)}}K\left[\frac{(a_2-a_3)(a_1-2x+1)}{(a_1-a_3)(a_2-2x+1)}\right], & x\in\left(\frac{1+a_2}{2},\frac{1+a_1}{2}\right], \\
        0, & x\not\in[\frac{1+a_3}{2},\frac{1+a_1}{2}],
    \end{cases}
\end{align}
where $K(\kappa)$ stands for the complete elliptic integral of the first kind,
$$K(\kappa)=\int_0^1 \frac{dt}{\sqrt{(1-t^2)(1-\kappa^2t^2)}}.$$

\section{Local amplitude-damping channels fidelity distribution calculation}\label{app:TwoADCCalculation}

The fidelity takes the form:
\begin{align}
    F_j(\vect)=\frac{1}{2}\left\{1+\frac{\pb[\mw_{j}]_{33}t_3+(t_1^2+t_2^2)\sqrt{(1-\pa)(1-\pb)}+t_3^2[(1-\pa)(1-\pb)+\pa\pb]}{1+\pa[\mw_j]_{33}t_3}\right\},
\end{align}
As from Eqs.~\eqref{eq:BellstatesCorrMat} we have $[\mw_1]_{33}=[\mw_2]_{33}=1$ and $[\mw_3]_{33}=[\mw_4]_{33}=-1$, we can change the index $j\in\{1,2,3,4\}$ by $k\in\{0,1\}$ according to $[\mw_j]_{33}\to (-1)^k$. Simplifying,
\begin{align}
    F_k(t_3)&=\frac{1}{2}\left[1+O_k(t_3)\right],\\
    O_k(t_3)&=\frac{(-1)^k\pb t_3+t_3^2g(\pa,\pb)+\sqrt{(1-\pa)(1-\pb)}}{1+\pa(-1)^kt_3},
\end{align}
where we have replaced: $1-t_1^2-t_2^2=t_3^2$. Thus, $F_k(\vect)$ does not depend on the azimuthal angle $\phi$. According to Appendix~\ref{app:Distr}, Eq.~\eqref{eq:probabilityDistrDelta},
\begin{align*}
   \pdf(F) &= \sum_k \pdf_k(F) \nonumber \\
   \pdf_k(F) &=\int_1^1 dt_3 \ \frac{[1+(-1)^kt_ 3\pa]}{4}\ \delta\!\left[F_k(t_3)-F\right],
\end{align*}
thus, the probability density function is characterized by the solutions of:
\begin{align}
    2F-1=O_k(t_3),
\end{align}
which we choose to label as $z_{kl}$ with $l\in\{0,1\}$. Note that, in this case, we can write:
\begin{align}
    \delta\!\left[F_k(t_3)-F\right] = \sum_l \frac{\delta(z_{kl}-t_3)}{|\left.\partial_{t_3}F_k(t_3)\right|_{t_3=z_{kl}}|}.
\end{align}
We have written all the main ingredients leading to Result~\ref{res:Distr}.

Let us consider now the fidelity given $k$, $F_k(\vect)$, optimized over $\vect\in S$, i.e.
\begin{align}
    F_{k}^{\min}(\pa,\pb) &= \min_{\vect\in S} F_k(\vect)\\
    F_{k}^{\max}(\pa,\pb) &= \max_{\vect\in S} F_k(\vect).
\end{align}
For each $k$, there are two critical points and two extremes characterizing, respectively, the maximal and minimal points which $F_k(\vect)$ may take. Fortunately, analyzing $F_0(\vect)$ or $F_1(\vect)$ leads to the same result, specifically, $F_{k}^{\max}=:F^{\max}$ and $F_{k}^{\min}=:F^{\min}$ are independent of $k$; these functions are plotted in Fig.~\ref{fig:ConditionalFidelityOpt}.

\end{document}